\documentclass[a4paper,12pt,abstract=true]{scrartcl}
\usepackage{authblk}
\usepackage{amsmath,amssymb}
\usepackage{bm}
\usepackage{graphicx,color}
\usepackage[hidelinks]{hyperref}

\usepackage[all,warning]{onlyamsmath}
\RequirePackage[l2tabu, orthodox]{nag}

\newbox{\ORCIDicon}
\sbox{\ORCIDicon}{\large
                  \includegraphics[width=0.8em]{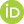}}

\begin{document}

\titlehead{\hfill OU-HET 1033, KYUSHU-HET-202}

\title{QED background against atomic neutrino process 
       with initial spatial phase}

\author[1]{Minoru~Tanaka\,\href{https://orcid.org/0000-0001-8190-2863}
                               {\usebox{\ORCIDicon}}
           \thanks{Email: \texttt{tanaka@phys.sci.osaka-u.ac.jp}}}

\affil[1]{Department of Physics, Graduate School of Science, 
          Osaka University, Toyonaka, Osaka 560-0043, Japan} 

\author[2]{Koji~Tsumura}
\affil[2]{Department of Physics, Kyushu University, 744 Motooka, Nishi-ku,
          Fukuoka, 819-0395, Japan}

\author[3]{Noboru~Sasao}
\affil[3]{Research Institute for Interdisciplinary Science, 
          Okayama University, Tsushima-naka 3-1-1, Kita-ku,
          Okayama 700-8530, Japan}

\author[3]{Satoshi~Uetake}

\author[3]{Motohiko~Yoshimura}

\date{\normalsize\today}

\maketitle

\begin{abstract}
Atomic deexcitation emitting a neutrino pair and a photon
is expected to provide a novel method of neutrino physics
if it is enhanced by quantum coherence in a macroscopic target.
However, the same enhancement mechanism 
may also lead to a serious problem of 
enhanced QED background process.
We show that the QED background can be suppressed enough
in the photonic crystal waveguide by using the spatial
phase that is imprinted in the process of initial coherence generation
in the target at excitation.
\end{abstract}

\newpage

\section{\label{Sec:Intro} Introduction}
\begin{figure}
 \centering
 \includegraphics[width=0.3\textwidth]{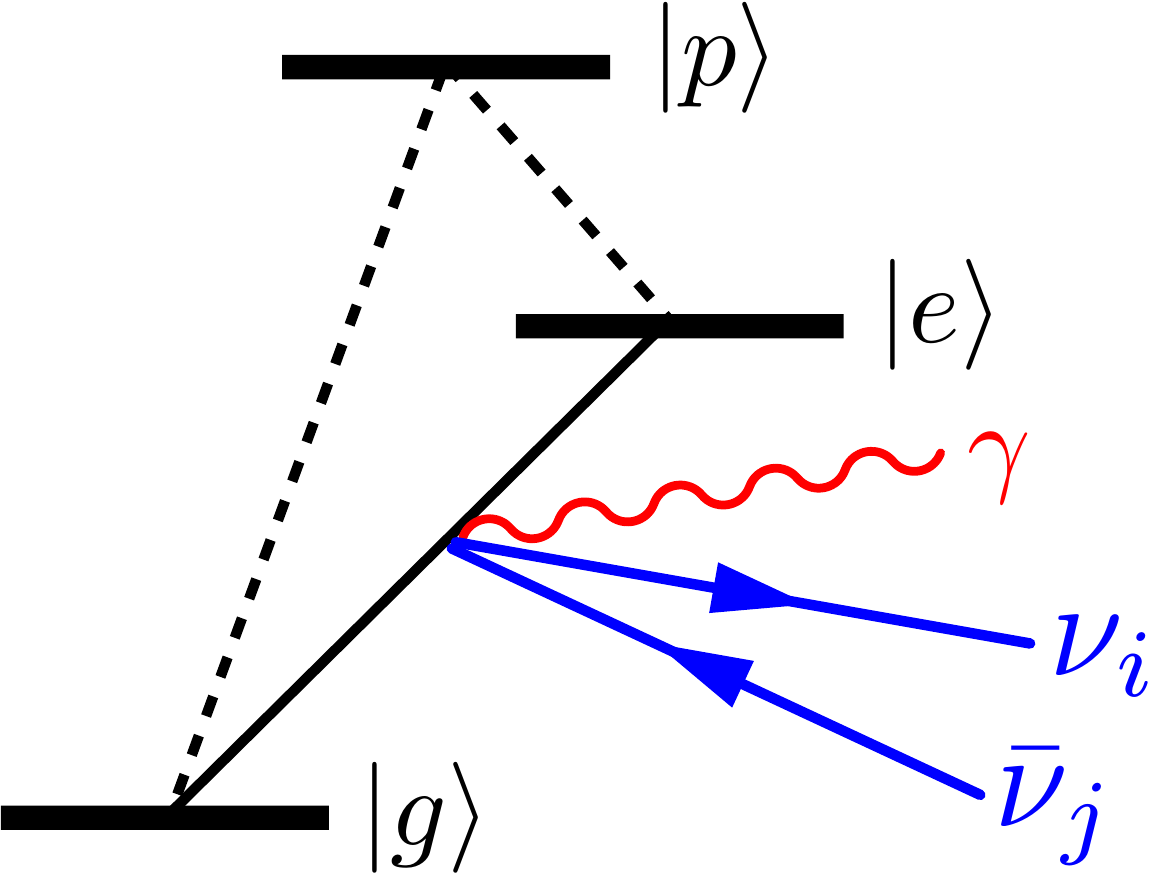}
 \caption{RENP scheme. A virtual intermediate state $|p\rangle$
          is shown as well as an excited initial state $|e\rangle$ 
          and the final ground state $|g\rangle$}.
 \label{Fig:RENPscheme}
\end{figure}

Neutrino pair emission in atomic transitions, 
which can produce nonrelativistic neutrinos, 
is conceived to further clarify neutrino properties.
Important questions to be addressed include the absolute 
values of neutrino masses and the discrimination of 
``Dirac or Majorana'' nature.
Radiative emission of neutrino pair (RENP),
$|e\rangle\to|g\rangle+\gamma+\nu_i+\bar\nu_j$,
has been considered as a candidate of such atomic neutrino
processes in Refs.~\cite{DinhPetcovSasaoTanakaYoshimura2012a,Fukumi2012a},
where $|e\rangle$ and $|g\rangle$ are a metastable excited state and
the ground state of an atom respectively, and $i$ and $j$ specify
neutrino mass eigenstates.
Three flavor scheme is assumed in the present work,
hence $i,j=1,2,3$.
The second order perturbation of QED and the weak four-fermion interaction
via a virtual intermediate state $|p\rangle$
leads to RENP process as depicted in Fig.~\ref{Fig:RENPscheme}.

RENP rate is quantitatively predictable in the standard
model of particle physics augmented by a mechanism of generating finite
neutrino masses and mixings and turns out to be strongly suppressed 
for an isolated atom owing to its extremely low energy scale ($\sim$ eV).
To overcome this small rate, 
an enhancement mechanism that makes use of 
quantum coherence in a macroscopic target~\cite{Yoshimura2008a},
called macrocoherence, is employed in the proposed experimental
scheme~\cite{DinhPetcovSasaoTanakaYoshimura2012a,Fukumi2012a}.

As a proof-of-concept of the macrocoherent amplification, 
the QED two-photon process in which a photon substitutes for 
the neutrino pair of RENP, $|e\rangle\to|g\rangle+\gamma+\gamma$,
has been studied both theoretically~\cite{YoshimuraSasaoTanaka2012a}
and experimentally%
~\cite{Miyamoto2014a,Miyamoto2015a,Miyamoto2017a,Hiraki2018a}.
In a series of experiments, 
the rate enhancement of $O(10^{18})$ was observed.

In addition to the remarkable rate amplification, the effective
momentum conservation, 
$\bm{p}_{eg}=\bm{p}_\gamma+\bm{p}_i+\bm{p}_j$, is the prominent
feature of the macrocoherence, 
where $\bm{p}_{\gamma,i,j}$ are 
momenta of the emitted particles and $\bm{p}_{eg}$ gives
the initial spatial phase (ISP) $e^{i\bm{p}_{eg}\cdot\bm{x}}$ 
in the macroscopic target created at excitation. 
We shall explain more on this.
We note that the momentum conservation, or atomic recoil, 
may be neglected to a good approximation in ordinary atomic 
radiation processes because of very small atomic recoil.
The macrocoherent momentum conservation holds 
irrespective of very small atomic recoil.
It corresponds to the phase-matching condition in 
the quantum electronics.

We can regard $\bm{p}_{eg}$ of the ISP as the momentum of 
the parent particle. 
Combined with the energy conservation, the RENP process
follows the same kinematics as in the three-body decay of
a particle of four-momentum $(E_{eg},\bm{p}_{eg})$, where
$E_{eg}$ is the level splitting between $|e\rangle$ and $|g\rangle$.
The four-momentum conservation results in the six thresholds in
the photon energy for possible neutrino pairs of $(i,j)$.
Hence, information on the neutrino masses can be obtained from
the photon spectrum in RENP. 
The threshold location in the RENP spectrum can be
searched by a frequency scan of the trigger laser. 

RENP process with a nonvanishing $\bm{p}_{eg}$, called boosted RENP, 
is studied in detail in Ref.~\cite{TanakaTsumuraSasaoUetakeYoshimura2017a}.
We can arrange the effective mass of the initial state,
$\sqrt{E_{eg}^2-\bm{p}_{eg}^2}$, at will
by varying $\bm{p}_{eg}$, so that
the effective range of neutrino masses in the RENP spectrum is varied.
In this work, we make use of $\bm{p}_{eg}$ to control 
the QED background of RENP.
 
\begin{figure}
 \centering
 \includegraphics[width=0.5\textwidth]{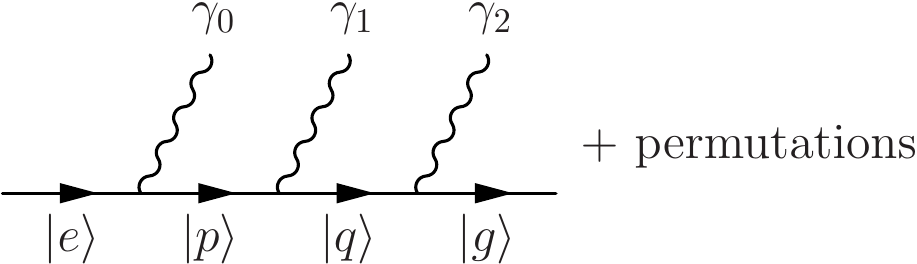}
 \caption{Feynman diagram of McQ3. Possible permutations of
          the photons are not explicitly shown.}
 \label{Fig:McQ3}
\end{figure}

The macrocoherent amplification also applies to QED processes of
multiphotons and leads to macrocoherent QED processes of $n$ photons
(McQ$n$)~\cite{YoshimuraSasaoTanaka2015a}. 
The most dangerous background is the three-photon process, McQ3,
$|e\rangle\to|g\rangle+\gamma_0+\gamma_1+\gamma_2$, where
$\gamma_0$ is the trigger photon and $\gamma_{1,2}$ are dynamical photons.
We present the Feynman diagram of McQ3 in Fig.~\ref{Fig:McQ3}. 
The neutrino pair in RENP is replaced by a pair of extra photons 
($\gamma_1+\gamma_2$) and thus their kinematics are similar.
The whole range of the invariant mass of 
the RENP neutrino pair is covered by that of the photon pair in McQ3.
It is shown that the rate of McQ3 is $O(10^{20})$ Hz,
while the amplified RENP rate is order of mHz for a xenon
transition~\cite{YoshimuraSasaoTanaka2015a}.

We examined the possibility to suppress McQ3 using the Bragg fiber%
~\cite{YehYarivMarom1978a},
a kind of hollow-core photonic crystal fibers~\cite{JJWM2nd}
that exhibit photonic band structures due to the periodicity
of refractive index~\cite{TanakaTsumuraSasaoYoshimura2017a}. 
Loading up the hollow core with the RENP target atoms, 
the McQ3 process becomes forbidden if one of extra photons
in McQ3 is emitted in the band gap of an ideal fiber.
RENP neutrinos are not affected by the fiber practically.
It has turned out that the Bragg fiber of the index contrast 
realized in the literature~\cite{Fink1999a} does not have 
a sufficiently wide band gap to cover the whole phase space of
extra photons in McQ3.

We assumed $\bm{p}_{eg}=0$ (no boost) in our previous work%
~\cite{TanakaTsumuraSasaoYoshimura2017a}.
In the present work, we study the effect of nonvanishing $\bm{p}_{eg}$
in McQ3 (boosted McQ3). In particular, we present a method 
to suppress McQ3 with the Bragg fiber of realistic index contrast
by locking the boost factor $\bm{p}_{eg}$ to the wave vector of
the trigger photon (boost-trigger lock, BTL).

The rest of the paper is organized as follows.
We briefly review the mechanism of macrocoherent amplification in 
Sec.~\ref{Sec:MC}.
The kinematics and rate of the boosted McQ3 in the free space
are given in Sec.~\ref{Sec:bMcQ3FS}.
Section~\ref{Sec:bMcQ3WG} describes the boosted McQ3 in the photonic
crystal waveguide as well as the feature of the Bragg fiber.
In Sec.~\ref{Sec:BTL}, we introduce the BTL scheme and 
present the numerical result of the rate suppression of the boosted 
McQ3 in the Bragg fiber.
The RENP spectrum in the BTL scheme is also shown in Sec.~\ref{Sec:BTL}.
Section~\ref{Sec:Conclusion} is devoted to our conclusion.

\section{\label{Sec:MC} Macrocoherent amplification mechanism}
Each atom in the macroscopic target of the RENP and its QED background 
processes may be regarded as the two-level quantum system of 
the ground state $|g\rangle$ and the excited state $|e\rangle$.
The deexcitation process emitting plural particles is described by 
the operator
$A:=\sum_a e^{-i\sum_i{\bm{p}_i}\cdot\bm{x}_a}|g\rangle_a\,{_a\!\langle e|}$,
where $\bm{x}_a$ is the position of the $a$th atom in the target and 
$i$ denotes the emitted particles. We have explicitly shown the phase
factor that represents the plane waves of the emitted particles and
left out coupling constants and other factors irrelevant in the following 
discussion in this section.

The deexcitation rate of the initial state $|\Psi\rangle$
is proportional to the squared wave function of the final state,
$|A|\Psi\rangle|^2=\langle\Psi|A^\dagger A|\Psi\rangle
 =\text{tr}A\hat\rho A^\dagger$,
where the density operator, $\hat\rho:=\Pi_a\hat\rho_a$, is introduced.
It is straightforward to obtain 
\begin{align}
\text{tr}A\hat\rho A^\dagger
=\sum_{a\neq a'}e^{i\sum_i{\bm{p}_i}\cdot(\bm{x}_{a'}-\bm{x}_a)}
 \langle e|\hat\rho_a|g\rangle \langle g|\hat\rho_{a'}|e\rangle
 +\sum_a\langle e|\hat\rho_a|e\rangle\,,
\end{align}
where the subscripts of the state vectors are omitted. 
The first sum is a double sum and represents the coherent contribution, 
while the second is single and incoherent. 
We neglect the latter provided that $|\langle e|\hat\rho_a|g\rangle|$
is sizable.
The coherence of the target is quantified by 
$|\langle e|\hat\rho_a|g\rangle|$.

Since a metastable state is favored as the upper state $|e\rangle$,
we assume that the ordinary electric dipole transition between
$|g\rangle$ and $|e\rangle$ is forbidden.
It turns out that the two-photon absorption process, 
$|g\rangle+\gamma+\gamma\to|e\rangle$,
is appropriate to prepare the initial state of the RENP with nonvanishing 
$\langle e|\hat\rho_a|g\rangle$. We note that this scheme of coherence
generation has already been realized for para-hydrogen~\cite{Hiraki2018a}.
As is described in Ref.~\cite{TanakaTsumuraSasaoUetakeYoshimura2017a},
the absorption of two photons of four-momenta $(\omega_1,\bm{k}_1)$
and $(\omega_2,\bm{k}_2)$ with $\omega_1+\omega_2=E_{eg}$ imprints
the ISP as 
$\langle e|\hat\rho_a|g\rangle=e^{i\bm{p}_{eg}\cdot\bm{x}_a}\rho_{a,eg}$,
where $\bm{p}_{eg}=\bm{k}_1+\bm{k}_2$.
In the slowly varying envelope approximation, in which the dependence
of $\rho_{a,eg}$ on $a$ is neglected in the leading order,
we obtain
\begin{align}
\text{tr}A\hat\rho A^\dagger\simeq
|\rho_{eg}|^2\sum_a e^{i(\bm{p}_{eg}-\sum_i{\bm{p}_i})\cdot\bm{x}_a}
             \sum_{a'} e^{-i(\bm{p}_{eg}-\sum_i{\bm{p}_i})\cdot\bm{x}_{a'}}\,.
\end{align}

Taking the limit of large number of atoms $N$ and 
large volume $V$ with the number density $n:=N/V$ fixed,
we find that the macrocoherently amplified rate is proportional to
\begin{align}\label{Eq:M2}
\text{tr}A\hat\rho A^\dagger\simeq
|\rho_{eg}|^2\frac{N^2}{V}(2\pi)^3\delta^3(\bm{p}_{eg}-\sum_i\bm{p}_i)\,.
\end{align}
The delta function in Eq.~\eqref{Eq:M2} combined with the one of 
the energy conservation, which is not explicitly shown, implies 
the energy-momentum conservation including the ISP in the macrocoherent
processes.

\section{\label{Sec:bMcQ3FS} Boosted McQ3 in the free space}

\subsection{\label{Sec:bMcQ3K} Kinematics} 
We consider the McQ3 process with the ISP,
$|e\rangle\to|g\rangle+\gamma_0(p_0)+\gamma_1(p_1)+\gamma_2(p_2)$,
where the four-momenta of the photons are denoted by
$p_i$ ($i=0,1,2$) and $\gamma_0$ is the trigger photon.
The four-momentum conservation owing to the macrocoherence is
expressed as
\begin{align}\label{Eq:4MC}
P^\mu=p_0^\mu+p_1^\mu+p_2^\mu\,,
\end{align}
where $P^\mu=(E_{eg},\bm{p}_{eg})$ and $p_i^\mu=(E_i,\bm{p}_i)$.
The two-photon absorption process of asymmetric antiparallel laser
irradiation is supposed to provide $\bm{p}_{eg}$ so that $P^2>0$.
We note that, in the relativistic three-body decay kinematics of 
Eq.~\eqref{Eq:4MC}, the invariant mass of the initial state must be
positive for the RENP process with a massive neutrino to take place%
~\cite{TanakaTsumuraSasaoUetakeYoshimura2017a}. 

It is convenient to introduce the four-momentum of the $\gamma_1\gamma_2$
system,
\begin{align}
q^\mu:=p_1^\mu+p_2^\mu=P^\mu-p_0^\mu\,,
\end{align}
where $q^2>0$ is also required for the RENP with 
the neutrino pair including at least one massive neutrino.
The temporal and spatial components are explicitly given by
\begin{align}\label{Eq:q}
(q^0,\bm{q})=(E_{eg}-E_0,\bm{p}_{eg}-\bm{p}_0)\,,\ q^0>0\,.
\end{align}
We obtain the following relation of the photon energy 
$E_{1,2}$ and the angle $\theta_{1,2}$
between $\bm{q}$ and the momentum of the each emitted 
photon by solving $(q-p_{1,2})^2=0$,
\begin{align}\label{Eq:costh}
\cos\theta_i=\frac{q^0}{|\bm{q}|}-\frac{q^2}{2|\bm{q}|E_i}\,,\ i=1,2\,,
\end{align}
or equivalently
\begin{align}
E_i=\frac{q^2}{2(q^0-|\bm{q}|\cos\theta_i)}\,,\ i=1,2\,.
\end{align}
The range of $E_i$ is determined by $|\cos\theta_i|\leq 1$,
\begin{align}
E_{i,\text{max}}&=\frac{1}{2}(q^0+|\bm{q}|)
                 =\frac{1}{2}(E_{eg}-E_0+|\bm{p}_{eg}-\bm{p}_0|)\,,
 \label{Eq:Emax}\\
E_{i,\text{min}}&=\frac{1}{2}(q^0-|\bm{q}|)
                 =\frac{1}{2}(E_{eg}-E_0-|\bm{p}_{eg}-\bm{p}_0|)\,.
 \label{Eq:Emin}
\end{align}

\subsection{Free-space rate}
The differential spectral rate of McQ3 with a boost $\bm{p}_{eg}$ 
in the free space is expressed as
\begin{align}\label{Eq:McQ3FS}
\frac{d\Gamma_\text{FS}}{dE_1}=\frac{\Gamma_0}{|\bm{q}|}|D|^2 E_1^2 E_2^2\,,
\end{align}
where $E_2=E_{eg}-E_0-E_1$.
The energy denominator factor $D$ including all possible permutations
in Fig.~\ref{Fig:McQ3} is defined by
\begin{align}
D:=&\phantom{+}\frac{1}{E_{pe}+E_0}
    \left(\frac{1}{E_{qg}-E_1}+\frac{1}{E_{qe}+E_0+E_1}\right)\nonumber\\
   &+\frac{1}{E_{pe}+E_1}
     \left(\frac{1}{E_{qg}-E_0}+\frac{1}{E_{qe}+E_0+E_1}\right)\nonumber\\
   &+\frac{1}{E_{pg}-E_0-E_1}
     \left(\frac{1}{E_{qg}-E_0}+\frac{1}{E_{qg}-E_1}\right)\,,
\end{align}
where $E_{pe}:=E_p-E_e$ with $E_{p(e)}$ the energy of 
$|p(e)\rangle$ and similar for the other combinations of states. 
The overall rate $\Gamma_0$ is given by
\begin{align}
\Gamma_0=\frac{3}{2}\pi^2n^2V
         \frac{\gamma_{gq}\gamma_{qp}\gamma_{pe}}
              {|E_{qg}|^3|E_{pq}|^3|E_{ep}|^3}
         |\mathcal{E}_\text{trig}|^2\,,
\end{align}
where $\mathcal{E}_\text{trig}$ denotes the average of the trigger electric
field in the target, $\gamma_{gq}$ is the $A$ coefficient of the
transition between $|g\rangle$ and $|q\rangle$ and so on.
Equation \eqref{Eq:McQ3FS} reproduces the McQ3 rate with no boot (no ISP)
in Ref.~\cite{YoshimuraSasaoTanaka2015a} taking $|\bm{q}|=E_0$.

\section{\label{Sec:bMcQ3WG} Boosted McQ3 in the photonic crystal waveguide}
It is desired to suppress the boosted McQ3 background by 20 orders or more
for observing the RENP process.
We have examined the background suppression mechanism by the photonic 
band gap in the photonic crystal waveguide in the case of no boost%
~\cite{YoshimuraSasaoTanaka2015a,TanakaTsumuraSasaoYoshimura2017a}.
In this section, we apply the same suppression mechanism to the boosted
McQ3. 

\subsection{\label{Sec:BF} Bragg fiber and its band structure}
\begin{figure}
 \centering
 \includegraphics[width=0.3\textwidth]{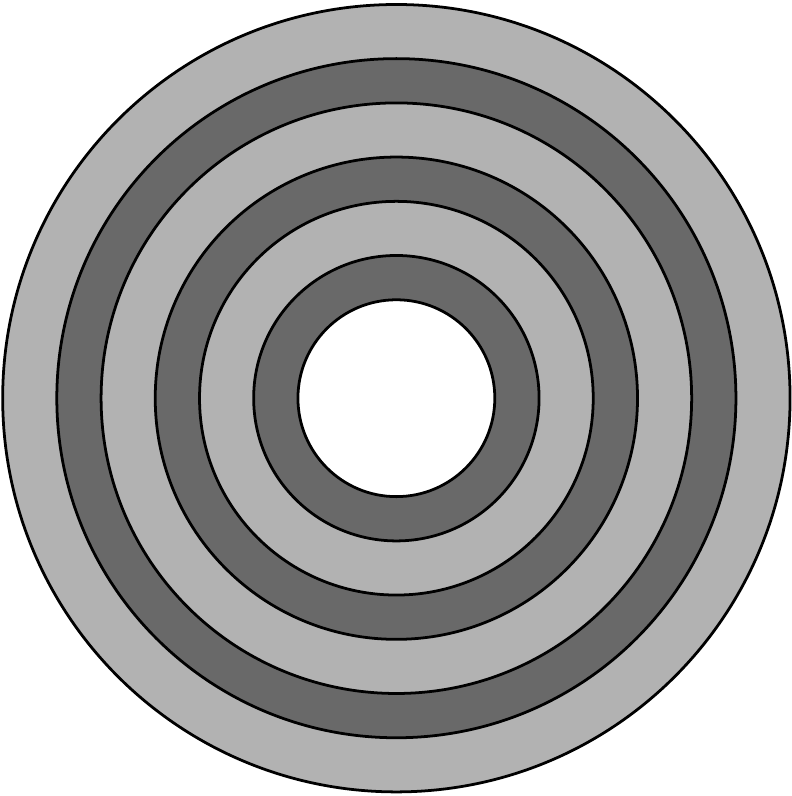}
 \caption{Cross section of the Bragg fiber. 
          The dark (light) gray layers have refractive index $n_1$ ($n_2$)
          and thickness $a_1$ ($a_2$).}
 \label{Fig:BF}
\end{figure}
We consider the Bragg fiber as an example of photonic crystal waveguides.
The cross section of the Bragg fiber is illustrated in Fig.~\ref{Fig:BF}.
It consists of the hollow core of radius $r_c$ and the dielectric 
cladding of $N_p$ layer pairs of alternating refractive indices, 
$n_1$ and $n_2$.
The $n_1$ and $n_2$ layer thicknesses are $a_1$ and $a_2$ respectively. 

The emission rate in a cavity like the core of the Bragg fiber
is described by the Purcell factor~\cite{Purcell1946a}, $F_P(E,\beta)$, 
which is a function of the energy $E$ of photon and the propagation
constant $\beta$. 
The latter is the projection of the wave vector onto the propagation 
(fiber) axis. We introduce 
\begin{align}\label{Eq:beta}
\beta_i:=E_i|\cos\theta_i|\,,\ i=1,2
\end{align}
for the emitted photons in McQ3, so that $\beta_i$ is non-negative.
This is because of the symmetry of the Bragg fiber under the flip of 
its axis direction, namely $F_P(E,\beta)=F_P(E,-\beta)$.

The Purcell factor is defined as the ratio of the emission power in
the cavity to that in the free space.
In the quantum theory of radiation, it is interpreted as the ratio of 
the state numbers of photon in the cavity and the free space. 
The calculation of the Purcell factor of the Bragg fiber is
described in Ref.~\cite{TanakaTsumuraSasaoYoshimura2017a} in detail.

\begin{figure}
 \centering
 \includegraphics[width=0.6\textwidth]{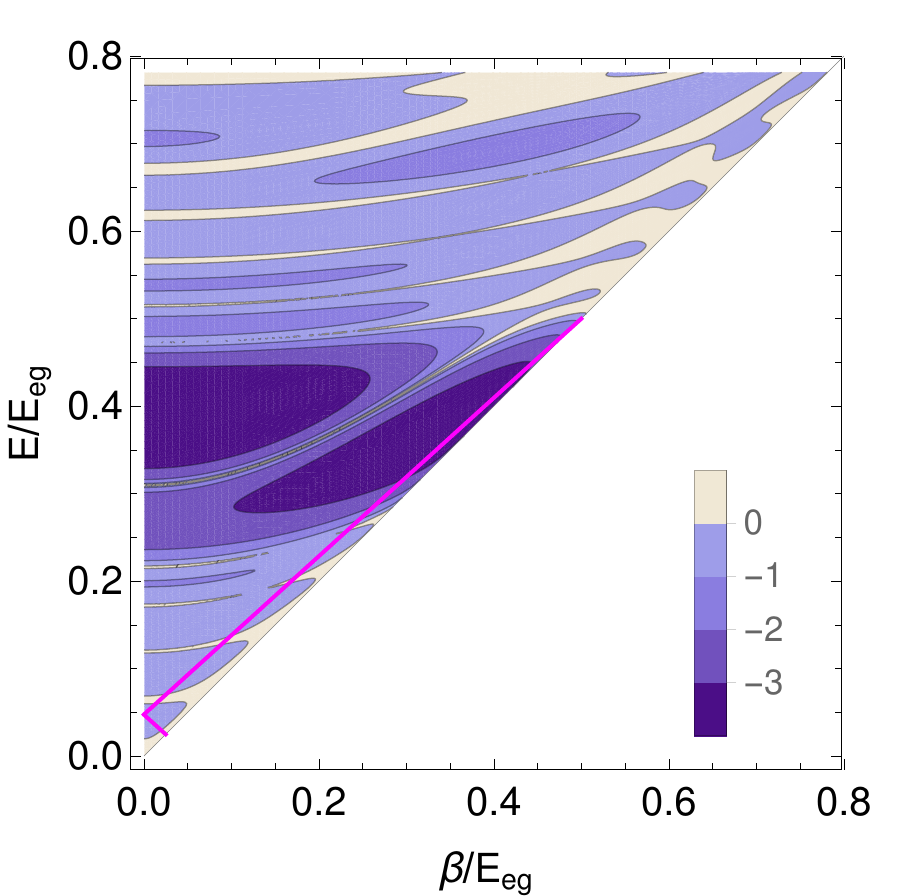}
 \caption{Purcell factor of the Bragg fiber of $(n_1,n_2)=(4.6,1.6)$.
          The scale is decimal logarithmic.
          The solid magenta line indicates the McQ3 physical line.}
 \label{Fig:FpNB}
\end{figure}

Figure~\ref{Fig:FpNB} illustrates the Purcell factor of the Bragg fiber 
that is designed to suppress the McQ3 with no boost. 
The relevant fiber parameters are
$(n_1,n_2)=(4.6,1.6)$, $a_2/a_1=\sqrt{n_1^2-1}/\sqrt{n_2^2-1}\simeq 3.6$, 
$r_c/(a_1+a_2)=2$ and $N_p=5$.
Darker regions represent stronger suppressions by smaller Purcell factors.
The photonic band gap is clearly seen around $E/E_{eg}\sim 0.4$.
The solid magenta line represents the physical
line on which the McQ3 photons locate in the case of no boost
and $E_0=0.95 E_{eg}/2$.
The expression of the physical line is easily obtained from
Eqs.~\eqref{Eq:costh} and \eqref{Eq:beta}, and the same for
the two emitted photons. 
In Fig.~\ref{Fig:FpNB}, the period of the cladding 
(in other words, the lattice spacing of the photonic crystal
that specifies the overall scale), $a_1+a_2$, 
is chosen to satisfy $E_{eg}(a_1+a_2)=0.64$ in order to realize
the maximal suppression of the McQ3 with no boost and 
$E_0=0.95 E_{eg}/2$ for the given set of the other fiber parameters.

\subsection{Rate in the photonic crystal waveguide}
We consider the McQ3 process with a boost in a photonic crystal waveguide
like the Bragg fiber. The directions of the boost momentum $\bm{p}_{eg}$ and
the trigger photon momentum $\bm{p}_0$ are taken to be the same as 
the direction of the propagation in the waveguide.
The rate is written in terms of the free-space one in Eq.~\eqref{Eq:McQ3FS}
and the Purcell factors $F_P(E,\beta)$ as
\begin{align}
\frac{d\Gamma_\text{WG}}{dE_1}=
 \frac{d\Gamma_\text{FS}}{dE_1}F_0 F_1 F_2\,,
\end{align}
where $F_i=F_P(E_i,\beta_i)$ ($i=0,1,2$), the propagation constant
$\beta_0$ is given by $\beta_0=E_0$ and
$\beta_{1,2}$ are defined in Eq.~\eqref{Eq:beta}.

As in the case of no boost~\cite{TanakaTsumuraSasaoYoshimura2017a}, 
the relative improvement of background suppression in the waveguide
is quantified by
\begin{equation}\label{Eq:rWGFS}
r_\text{WG/FS}:=\frac{1}{\Gamma_\text{FS}(\bm{p}_{eg},E_0)}
                \int\frac{d\Gamma_\text{FS}}{dE_1}F_1F_2dE_1\,.
\end{equation}
We note that the Purcell factor of the trigger $F_0$ disappears
because it is common for the signal (RENP) and the background (McQ3).
In order to obtain a sufficient suppression, it is required that 
at least one of the emitted photons $\gamma_1$ and $\gamma_2$
is in the band gap of the photonic crystal so that $F_1F_2$ is
tiny in the whole phase space.

\section{\label{Sec:BTL} Method of boost-trigger lock (BTL)}
\subsection{Problem in the case of no boost}
It is shown in our previous work~\cite{TanakaTsumuraSasaoYoshimura2017a}, 
the band gap of the Bragg fiber of $(n_1, n_2)=(4.6, 1.6)$, which is
fabricated in the laboratory~\cite{Fink1999a}, is not wide enough to
suppress the McQ3 rate in the case of no boost. 
This is due to the rather wide photon spectrum,
$E_{i,\text{max}}=E_{eg}/2$ and $E_{i,\text{min}}=E_{eg}/2-E_0$,
as seen in Eqs.\eqref{Eq:Emax} and \eqref{Eq:Emin} with $\bm{p}_{eg}=0$.
The trigger photon energy $E_0$ must be close to $E_{eg}/2$ to probe 
the neutrino thresholds in RENP, which is give by 
$E_{eg}/2-(m_i+m_j)^2/2E_{eg}$ for $\bm{p}_{eg}=0$,
where $m_{i,j}$ denote the neutrino masses.
Hence, $0\lesssim E_i\leq E_{eg}/2$, as exemplified by 
the solid magenta line in Fig.~\ref{Fig:FpNB}.
The upper half of this range must be in the forbidden band near
the light line, defined by $E=\beta$, 
and this is not the case for the realistic
Bragg fiber as illustrated in Fig.~\ref{Fig:FpNB}.
The McQ3 suppression is $O(10^{-2})$ even for large $N_p$.
The detailed analysis of this difficulty is given in 
Ref.~\cite{TanakaTsumuraSasaoYoshimura2017a}.

\subsection{Boost-Trigger lock}
There are two possible directions to solve the above problem:
To make the band gap wider and/or to make the photon spectrum narrower. 
The former is investigated in our previous work%
~\cite{TanakaTsumuraSasaoYoshimura2017a}.
We can take the latter in the case of the boosted McQ3.

\begin{figure}[t]
 \centering
 \includegraphics[width=0.6\textwidth]{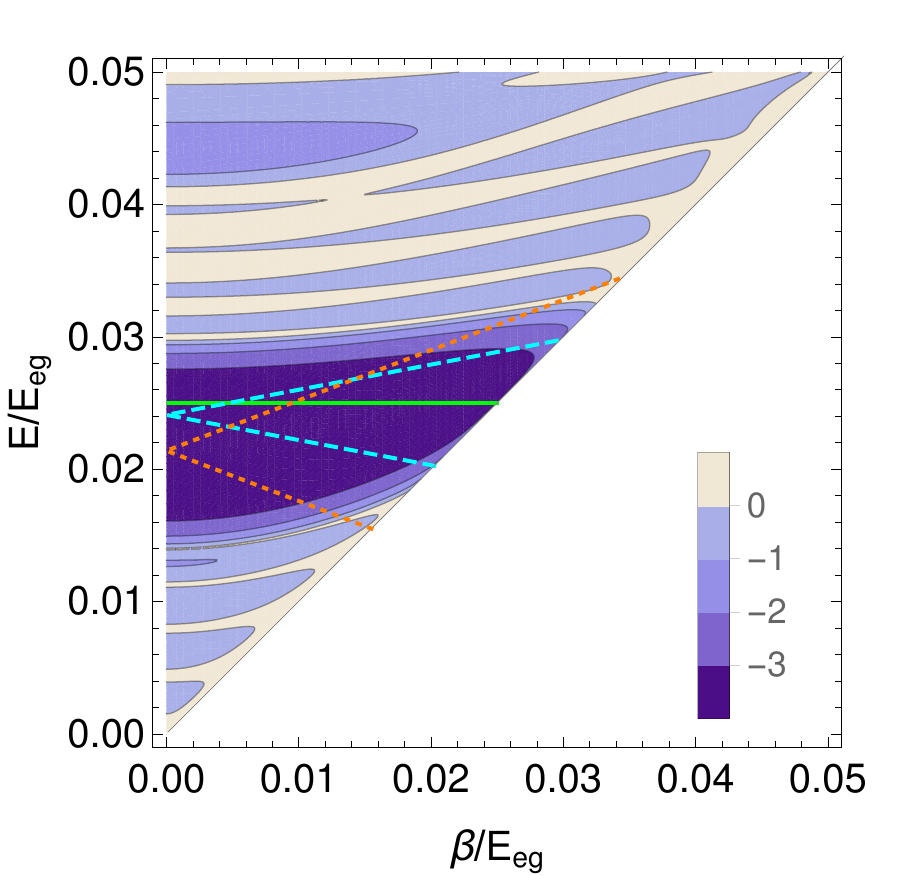}
 \caption{Purcell factor of the Bragg fiber designed to suppress 
          the McQ3 with the BTL method. 
          The index pair is the same as Fig.~\ref{Fig:FpNB}.
          The solid green horizontal line represents the McQ3 
          physical line in the BTL mode. 
          The cyan dashed and orange dotted lines express
          the McQ3 physical lines of off-BTL cases.}
 \label{Fig:FpBTL}
\end{figure}

Suppose that the trigger momentum is locked to that of the boost,%
\footnote{This may be achieved using the difference of two excitation
 lasers as the trigger.}
\begin{align}
\bm{p}_{eg}=\bm{p}_0\,,
\end{align}
so that $\bm{q}=0$, that is, the center of mass of the photon pair 
$\gamma_1\gamma_2$ is at rest in the laboratory frame.
Then, we find 
\begin{align}\label{Eq:EBTL}
E_{i,\text{max}}=E_{i,\text{min}}=\frac{q^0}{2}=\frac{1}{2}(E_{eg}-E_0)
=:E_\text{BTL}\,,
\end{align}
namely, the spectrum of the emitted photons is monochromatic for
a given trigger energy, $E_1=E_2=E_\text{BTL}$.
The propagation constants of the emitted photons are 
\begin{align}
\beta_i=E_\text{BTL}|\cos\theta_i|\,,\ i=1,2.
\end{align}
The physical line on which the McQ3 events lie in the $\beta$--$E$ plane
is the straight section defined by $0<\beta<E_\text{BTL}$ and
$E=E_\text{BTL}$.

In Fig.~\ref{Fig:FpBTL}, we illustrate the BTL physical 
line with $|\bm{p}_{eg}|=0.95 E_{eg}$, which is favored to enhance 
the signal sensitivity in the RENP of ytterbium target%
~\cite{TanakaTsumuraSasaoUetakeYoshimura2017a},
as well as the Purcell factor of the Bragg fiber appropriate to suppress
the McQ3 background with the BTL method.
The parameters of the Bragg fiber are the same as Fig.~\ref{Fig:FpNB}
except $r_c/(a_1+a_2)=1$ and $E_{eg}(a_1+a_2)=10$.
The BTL condition $|\bm{p}_{eg}|=E_0$ leads to the monochromatic
photons of $E_\text{BTL}=0.05 E_{eg}/2$.
We observe that the both the emitted McQ3 photons are always in
the forbidden band.
As mentioned above, the lattice spacing $a_1+a_2$ in 
Fig.~\ref{Fig:FpBTL}is chosen as $E_{eg}(a_1+a_2)=10$ and much larger 
than that in Fig.~\ref{Fig:FpNB}. 
This is because of the lower energy scale of the emitted photons
in the boosted McQ3 than the McQ3 without boost.

We also show two physical lines that deviate from the BTL condition
for comparison. 
The case of $|\bm{p}_{eg}|=0.99(0.98) E_0$ is expressed by
the cyan dashed (orange dotted) line.
We observe that the photon spectra extend toward
the outside of the bad gap in these cases.
We note, however, that the boost-trigger lock would not necessarily be exact.
A tiny deviation of $\bm{q}=\bm{p}_{eg}-\bm{p}_0$ from 0 only leads to
the energy width $E_{i,\text{max}}-E_{i,\text{min}}=|\bm{q}|$
as seen in Eqs.~\eqref{Eq:Emax} and \eqref{Eq:Emin}. 
The band gap that can accommodate this tiny energy width is
possible with a reasonable index contrast.

\subsection{Suppression of the BTL McQ3 in the Bragg fiber}
\begin{figure}
 \centering
 \includegraphics[width=0.6\textwidth]{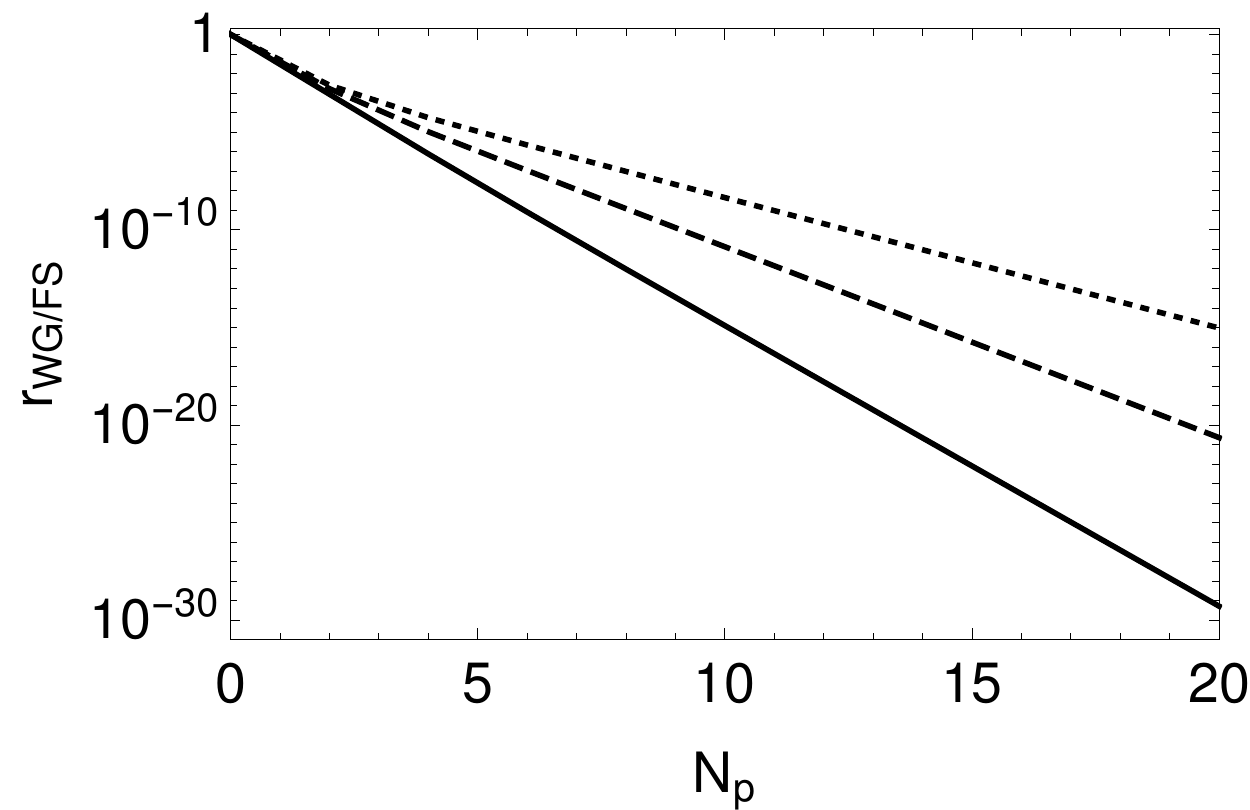}
 \caption{McQ3 suppression factors in the Bragg fiber as functions of
          the number of layer pairs $N_p$.
          The other fiber parameters are the same as Fig.~\ref{Fig:FpBTL}.
          The trigger energy is also the same as Fig.~\ref{Fig:FpBTL},
          $E_0=0.95 E_{eg}$. 
          Solid: the exact BTL case, $|\bm{p}_{eg}|=E_0$.
          Dashed (Dotted): the off-BTL case of 
          $|\bm{p}_{eg}|=0.990(0.988)E_0$.}
 \label{Fig:rBFFSBTLnp}
\end{figure}

The suppression factor of McQ3 in the Bragg fiber is given by 
Eq.~\eqref{Eq:rWGFS} with the Purcell factor of the Bragg fiber.
The solid line in Fig.~\ref{Fig:rBFFSBTLnp}
shows the McQ3 suppression factor in Eq.~\eqref{Eq:rWGFS} 
in the BTL mode with $E_0=0.95 E_{eg}$
as a function of the number of layer pairs $N_p$.
The other fiber parameters are the same as in Fig.~\ref{Fig:FpBTL}.
We observe the exponential behavior as inferred from the analytic
argument in Ref.~\cite{TanakaTsumuraSasaoYoshimura2017a}.
A suppression $10^{-23}$ or better is obtained for $N_p\gtrsim 16$.

In Fig.~\ref{Fig:rBFFSBTLnp}, we also present the McQ3 suppression
factors in two cases of off-BTL, 
$|\bm{p}_{eg}|=0.990(0.988) E_0$ in the dashed (dotted) line.
In the present setup, the 1\% deviation from the BTL limit is acceptable,
but the further deviation easily leads to the failure of suppression.

\begin{figure}
 \centering
 \includegraphics[width=0.6\textwidth]{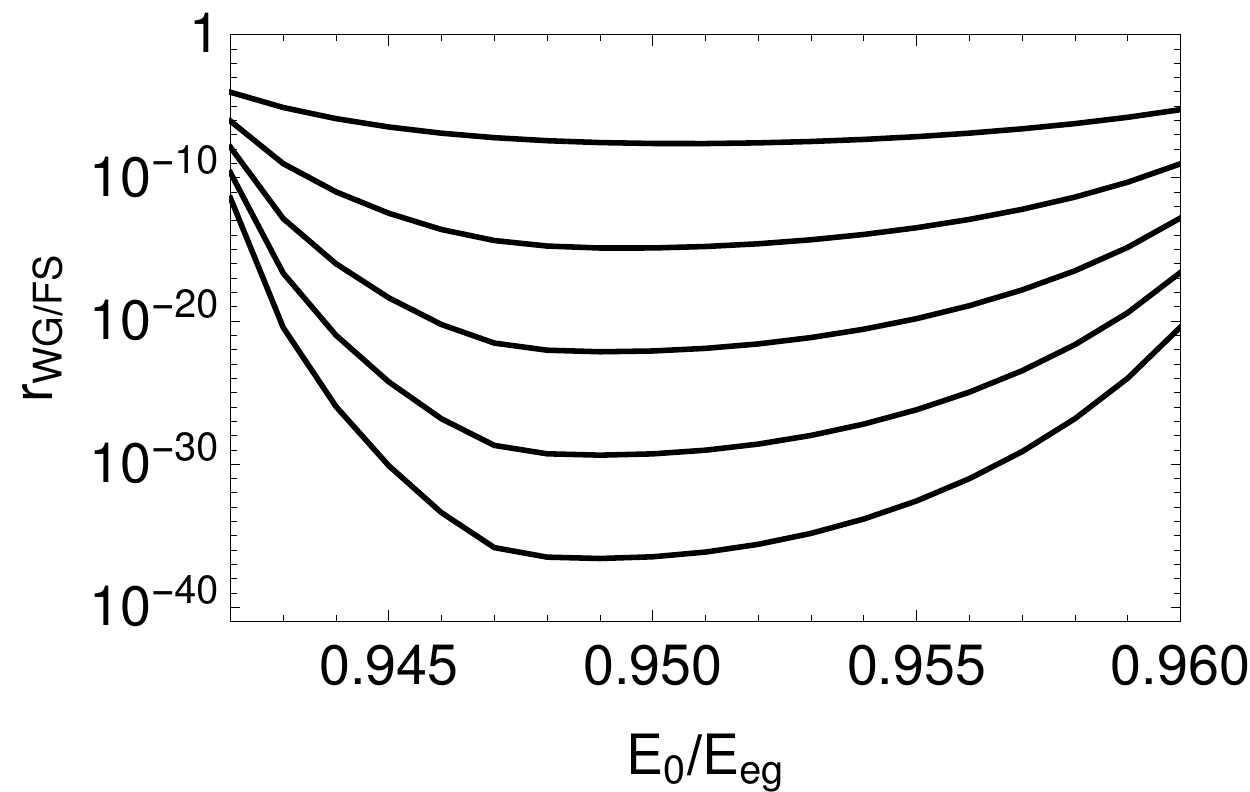}
 \caption{McQ3 suppression factors in the Bragg fiber with the BTL mode
          as functions of trigger energy for $N_p=5,10,15,20,25$ 
          (from top to bottom).
          The other fiber parameters are the same as in 
          Fig.~\ref{Fig:FpBTL}.}
 \label{Fig:rBFFSBTLnpom}
\end{figure}

In Fig.~\ref{Fig:rBFFSBTLnpom}, the McQ3 suppression factors in 
the Bragg fiber with the BTL mode as functions of trigger energy for 
$N_p=5,10,15,20,25$.
The other fiber parameters are the same as in Fig.~\ref{Fig:FpBTL}.
We observe e.g. that the suppression is sufficient in the trigger 
range of $0.945 E_{eg}<E_0< 0.955 E_{eg}$ for $N_p=20$. 
For the Yb case ($E_{eg}=2.14349$ eV) examined in 
Ref.~\cite{TanakaTsumuraSasaoUetakeYoshimura2017a}, 
this means that we can scan the trigger range between 
2.0256 eV and 2.0470 eV with suppressed McQ3 backgrounds. 
It is possible to modify this range by changing the overall scale
(lattice spacing, $a_1+a_2$) of the Bragg fiber. 
The width of this range is $\sim 20$ meV and practically independent
of this scaling since we are interested in the narrow region near
the endpoint ($\sim E_{eg}$, see below).

\subsection{RENP spectrum in the BTL mode}
As described above, the three momentum of the $\gamma_1\gamma_2$ system
vanishes in the BTL mode, namely $\bm{q}=0$. 
In other words, the $\gamma_1\gamma_2$ system is at rest in 
the laboratory frame.
This also applies to the case of RENP, in which the $\nu_i\bar\nu_j$ 
system has $\bm{q}=0$ and is at rest. 
Then the RENP spectral rate is simplified as
\begin{align}
\Gamma^{\nu\bar\nu}_\text{BTL}(E_0)=
 \Gamma^{\nu\bar\nu}_0\sum_{i,j}
 \frac{\beta_{ji}q^2}{6(E_{pg}-E_0)^2}\frac{E_0}{E_{eg}}
  \biggl[&|c^A_{ji}|^2\left\{2-\frac{m_j^2+m_i^2}{q^2}
                              -\frac{(m_j^2-m_i^2)^2}{q^4}\right\}\\
         &-6\delta_M\text{Re}(C^{A2}_{ji})\frac{m_jm_i}{q^2}\biggr]\,,
\end{align}
where $c^A_{ji}=U_{ej}^*U_{ei}-\delta_{ji}/2$ with $U$ being the PMNS 
matrix~\cite{PDG2018}, $q^2$ is the invariant mass of the neutrino pair
with $q^\mu$ given by Eq.~\eqref{Eq:q}, $\delta_M=0(1)$ for Dirac 
(Majorana) neutrinos, and
\begin{align} 
\beta_{ji}:=1-2\frac{m_j^2+m_i^2}{q^2}+\frac{(m_j^2-m_i^2)^2}{q^4}\,.
\end{align}
The overall rate $\Gamma^{\nu\bar\nu}_0$ is given in 
Ref.~\cite{TanakaTsumuraSasaoUetakeYoshimura2017a}. 

\begin{figure}
 \centering
 \includegraphics[width=0.6\textwidth]{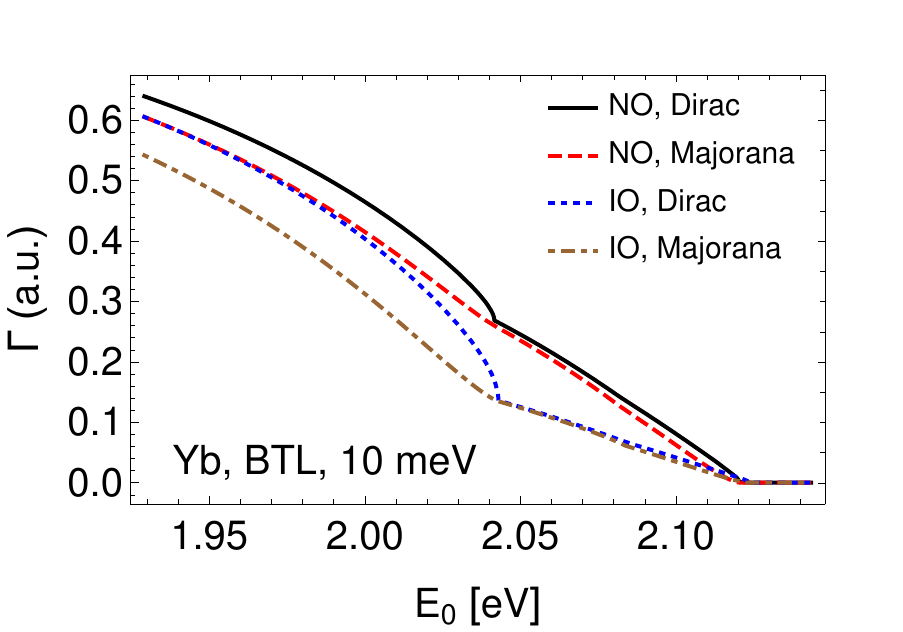}
 \caption{RENP spectrum in the BTL mode. The target is Yb and 
          the light neutrino mass is $10$ meV.
          The Majorana phases are taken to be 0.
          The same transition is employed as in 
          Ref.~\cite{TanakaTsumuraSasaoUetakeYoshimura2017a}. 
          The other neutrino parameters are the best-fit values 
          taken from NuFIT 2018~\cite{Esteban2017a}.}
 \label{Fig:YbBTL10}
\end{figure}

We stress that the difference between Dirac and Majorana neutrinos
and the dependence on the neutrino masses $m_{i,j}$ remain in 
the BTL mode.
Namely, the neutrino mass spectroscopy including
the Dirac-Majorana discrimination is possible in the BTL mode of RENP.
In Fig.~\ref{Fig:YbBTL10}, we illustrate the RENP spectrum in
the BTL mode for the case of Yb target ($E_{eg}=2.14349$ eV) and 
the lightest neutrino mass of 10 meV. 
See Ref.~\cite{TanakaTsumuraSasaoUetakeYoshimura2017a} for details
of the employed transition.
The kink structure at thresholds and the difference between Dirac and
Majorana are apparent.

The threshold of the trigger photon energy to produce 
a $\nu_i\bar\nu_j$ pair is given by $E_{eg}-(m_i+m_j)$
under the BTL condition.
Using the 2018 results of NuFIT~\cite{Esteban2017a} and 
Planck~\cite{Planck2018VI} collaborations, we find that
the upper bound of the lightest neutrino mass is 30.2 meV for
the normal ordering (NO) and 16.2 meV for the inverted ordering (IO).
The mass range of the heaviest neutrino is 
[49.9, 58.4]([49.6, 52.2]) meV for NO(IO).
It is possible to cover these mass ranges in the neutrino mass 
spectroscopy with a few fibers of different lattice spacings,
since each Bragg fiber of fixed lattice spacing covers the mass range
about 20 meV near the endpoint $E_{eg}$ corresponding
to the massless neutrinos as explained above.

\section{\label{Sec:Conclusion} Conclusion}
We have studied the QED background process McQ3 against the atomic
neutrino process RENP in the presence of the initial spatial phase.
The macrocoherence amplifies RENP, but at the same time amplifying 
QED backgrounds.

To suppress the McQ3 process, we consider the experimental scheme
that makes use of the hollow-core photonic crystal fiber.
The photonic band gap of the crystal prohibits the photon emission
in principle.
The width of the band gap must be wide enough so that the relevant 
energy range of the emitted photons in McQ3 is covered.

We have found that the energy spectrum of McQ3 can be controlled
by making the initial spatial phase factor 
$\bm{p}_{eg}$ locked to the momentum of 
the trigger photon (boost-trigger lock, BTL).
The four-momentum conservation dictated by the macrocoherence
implies that the center of mass of the emitted photon pair in McQ3
is at rest in the BTL mode, so that the photon spectrum is monochromatic.
We have shown that the sufficient suppression of McQ3 in the Bragg fiber
is possible using the BTL method. 

The RENP spectrum in the BTL mode has also been examined.
It is found that the dependence on the neutrino masses 
and the difference between Dirac and Majorana neutrinos
remain sizable in the RENP spectrum in the BTL mode.
We have found that a set of a few fibers of different
lattice spacings is sufficient to scan the region sensitive to
neutrino masses.

To summarize, the McQ3 QED background problem in the atomic neutrino
process RENP can be solved in principle using a photonic crystal fiber. 
The BTL method that makes use of the initial spatial phase works
with the Bragg fiber of the realistic combination of refractive indices.

\section*{Acknowledgments}
This work is supported in part by JSPS KAKENHI Grant Numbers 
JP 16H03993 (MT), 18K03621 (MT), 18H05543 (KT), 17H02895 (MY), 
15H02093 (NS) and 15K13486 (NS).

\providecommand{\href}[2]{#2}\begingroup\raggedright\endgroup

\end{document}